\documentclass{article}
\usepackage{slashed,epsfig,amsmath,amssymb,enumitem}
\usepackage[papersize={8.5in,11in}]{geometry}
\usepackage{graphicx}
\newcommand{\be}{\begin{equation}}
\newcommand{\ee}{\end{equation}}
\begin{document}
\title{Neutrino Masses and Helicities}
\author{J. W. Moffat\\~\\
Perimeter Institute for Theoretical Physics, Waterloo, Ontario N2L 2Y5, Canada\\
and\\
Department of Physics and Astronomy, University of Waterloo, Waterloo,\\
Ontario N2L 3G1, Canada}
\maketitle

%\date{\today}

%\thanks{PACS: 98.80.C; 04.20.G; 04.40.-b}

% ----------------------------------------------------------------

\begin{abstract}
Models of neutrinos and their mass generation by self-energy radiative corrections are formulated in an ultraviolet complete quantum field theory (UCQFT). A model of the three flavors of neutrinos as Majorana fermions is developed as the minimal model. A model incorporating an SU(2) singlet sterile neutrino can also be formulated.
\end{abstract}
\maketitle

\section{Introduction}

The standard model (SM)~\cite{Halzen,Peskin,Burgess,PDG,Higgs1964,Brout1964,Kibble1964,Higgs1966,Kibble1967} is based on the requirement that all interactions of particles are renormalizable. This imposes the constraint that the Lagrangians for the strong QCD, electromagnetic and weak interactions are gauge invariant according to the representations of the group $SU(3)_C\times SU(2)_L\times U(1)_Y$. In particular, the constraint that the SM model satisfy $SU(2)_L\times U(1)_Y$ gauge invariance and requiring the separate conservation of the three lepton numbers $L_e, L_\mu$ and $L_\tau$, predicts that in the SM the stable neutrinos, $\nu_e, \nu_\mu, \nu_\tau$ are exactly massless, with the left-handed $\nu_L$ together with its corresponding charged lepton, $\l^-$, only participating in charged current weak interactions. Experiments involving neutrinos demonstrate that they are massive and that they exchange flavors through neutrino oscillations. This indicates that the lepton numbers $L_e, L_\mu, L_\tau$ are not separately conserved. The experimental absence of a right-handed $SU(2)$ singlet neutrino and the need that the SM is renormalizable, together with the $SU(2)_L\times U(1)_Y$ gauge invariance constraint, excludes the possibility that neutrinos obtain their mass from the Higgs field. This means that the SM is an incomplete theory and needs to be modified. In particular, the infinite renormalizability requirement breaks down. This motivates us to formulate an alternative model based on an ultraviolet complete quantum field theory in which all QFT calculations are finite, Poincar\'e invariant and unitary to all orders of perturbation theory.

\section{Formulation of Alternative Model}

An alternative solution of the generation of particle masses is based on an ultraviolet complete quantum field theory (UCQFT)~\cite{Moffat1989,Moffat1990,MoffatWoodard1991,Moffat1991,WoodardKleppe1991,WoodardKleppe1992,Cornish1992a,Cornish1992b,
Cornish1992c,WoodardKleppe1993,ClaytonDemopolous1994,MoffatToth2010,Moffat2020,Moffat2021,GreenMoffat}. The UCQFT realizes a Poincar\'e invariant, unitary QFT and all quantum loop graphs are ultraviolet finite to all orders of perturbation theory. Although the field operators and the interactions of particles are nonlocal, the model satisfies microscopic causality~\cite{Moffat2020,Moffat2021}. The model is based on the experimentally known twelve flavor fermions and four vector bosons and the spin 0 scalar Higgs particle. However, for the neutrino sector, it is possible to hypothesize the existence of a sterile fourth isosinglet neutrino with zero quantum numbers that does not interact with matter.

The dynamics of the quarks and gluons are controlled by the quantum chromodynamics (QCD) Lagrangian:
\be
{\cal L}_{QCD}=\bar\psi_{qi}i\gamma^\mu(D_\mu)_{ij}\psi_{qj} - \frac{1}{4}G^a_{\mu\nu} G^{\mu\nu}_a,
\ee
where $\psi_{qi}$ is the quark field in the fundamental representation of the $SU(3)$ gauge group and $i$ and $j$ run from 1 to 3. The $D_\mu$ is the gauge covariant derivative:
\be
D_{\mu_{ij}}=\partial_\mu \delta_{ij} - i{\tilde g}_s(T_a)_{ij} A^a_\mu,
\ee
and
\be
G^a_{\mu\nu}=\partial_\mu A^a_\nu - \partial_\nu A^a_\mu + {\tilde g}_sf^{abc} A^b_\mu A^c_\nu,
\ee
where $A^a_\mu$ are the massless gluon fields.

The Lagrangian in the new alternative model for the electroweak sector is given by
\be
{\cal L}_{EW}={\cal L}_{\rm SMEW}+{\cal L}_{Mb} + {\cal L}_{mql} + {\cal L}_{\nu},
\ee
where ${\cal L}_{SMEW}$ is the SM gauge invariant electroweak Lagrangian:
\begin{align}
\label{Lagrangian}
{\cal L}_{\rm SMEW}==\sum_{\psi_L}\bar\psi_L\biggl[\gamma^\mu\biggl(i\partial_\mu - \frac{1}{2}{\tilde g}\tau^aW^a_\mu - {\tilde g}'\frac{Y}{2}B_\mu\biggr)\biggr]\psi_L
+\sum_{\psi_R}\bar\psi_R\biggl[\gamma^\mu\biggl(i\partial_\mu - {\tilde g}'\frac{Y}{2}B_\mu\biggr)\biggr]\psi_R\nonumber\\
 -\frac{1}{4}B^{\mu\nu}B_{\mu\nu}
-\frac{1}{4}W_{\mu\nu}^aW^{a\mu\nu}.
\end{align}
Here, the ${\tau}'s$ are the usual Pauli spin matrices and $\psi_L$ denotes a left-handed fermion (lepton or quark) doublet, and the $\psi_R$ denotes a right-handed fermion singlet. The fermion fields (leptons and quarks) have been written as $SU_L(2)$ doublets and U(1)$_Y$ singlets, and we have suppressed the fermion generation indices. We have $\psi_{L,R}=P_{L,R}\psi$, where $P_{L,R}=\frac{1}{2}(1\mp\gamma_5)$. Moreover, we have
\begin{equation}
\label{Bequation}
B_{\mu\nu}=\partial_\mu B_\nu-\partial_\nu B_\mu,
\end{equation}
\begin{equation}
W^a_{\mu\nu}=\partial_\mu W_\nu^a-\partial_\nu W_\mu^a-{\tilde g}f^{abc}W_\mu^bW_\nu^c.
\end{equation}
The vector boson mass Lagrangian is given by
\be
\label{bosemass}
{\cal L}_{Mb} = \frac{1}{2}M^2_WW^{a\mu} W^a_\mu + \frac{1}{2}M^2_BB^\mu B_\mu,
\ee
and the quark and charged lepton mass Lagrangian is
\be
\label{quarkleptonmass}
{\cal L}_{mql}=-\sum_{\psi_L^i,\psi_R^j}m_{ij}^f(\bar\psi_L^i\psi_R^j + \bar\psi_R^i\psi_L^j),
\ee
where $M_W$, $M_B$ and $m_{ij}^f$ denote the boson and the quark and charged lepton fermion masses, respectively. The ${\cal L}_\nu$ denotes the neutrino mass term. In the following, we will determine the form of ${\cal L}_\nu$.

The Lagrangian for the scalar Higgs boson is given by
\be
{\cal L}_{\rm Higgs}=\biggl|\biggr(i\partial_\mu -\frac{1}{2}{\tilde g_H}\tau^a W_\mu^a-{\tilde g_H}'\frac{Y}{2}B_\mu\biggr)\phi\biggr|^2
+\frac{1}{2}m_H^2\phi^2,
\ee
where $\phi$ is the isoscalar Higgs field. The photon-fermion Lagrangian is
\be
L_{\rm QED}=\sum_{\psi_L}\bar\psi_L\biggl[\gamma^\mu\biggl(i\partial_\mu - \frac{1}{2}{\tilde e}\biggr)A_\mu\biggr]\psi_L
+\sum_{\psi_R}\bar\psi_R\biggl[\gamma^\mu\biggl(i\partial_\mu - \frac{1}{2}{\tilde e}\biggr)A_\mu\biggr]\psi_R -\frac{1}{4}F^{\mu\nu}F_{\mu\nu}+{\cal L}_{m_f},
\ee
where ${\cal L}_{mf}$ denotes the fermion mass Lagrangian. Moreover,
\be
F_{\mu\nu}=\partial_\mu A_\nu-\partial_\nu A_\mu.
\ee

The fermion and boson fields are local fields and we have
\be
{\tilde g}_s=g_s{\cal E}(p^2/\Lambda_i^2),
\ee
\be
{\tilde g}(p^2)=g{\cal E}(p^2/\Lambda_i^2),
\ee
\be
{\tilde g}'(p^2)=g'{\cal E}(p^2/\Lambda_i^2),
\ee
\be
{\tilde e}(p^2)=e{\cal E}(p^2/\Lambda_i^2).
\ee
\be
{\tilde g_H}(p^2)=g_H{\cal E}(p^2/\Lambda_H^2),
\ee
\be
{\tilde g'_H}(p^2)=g'_H{\cal E}(p^2/\Lambda_H^2).
\ee

We define the entire function distribution operator ${\cal E}$ in terms of the kinetic operator ${\cal K}$:
\be
\label{Edistr}
{\cal E}=\exp\bigg(\frac{{\cal K}}{2\Lambda_i^2}\biggr).
\ee
The Feynman rules for the UCQFT follow as extensions of the local standard QFT. Every internal line in a Feynman diagram can be connected to a regulated propagator:
\be
\label{propagatorReg}
i\tilde\Delta=\frac{i{\cal E}^2}{{\cal K}}=i\int\frac{d\tau}{\Lambda_i^2}\exp\biggl(\tau\frac{\cal K}{\Lambda_i^2}\biggr),
\ee
where we have used the Schwinger proper time method to determine the propagator.

An additional auxiliary propagator was introduced in the formulation of finite QED~\cite{MoffatWoodard1991}:
\be
\label{auxpropagator}
-i\hat\Delta=\frac{i(1-{\cal E}^2)}{{\cal K}}=-i\int_0^1\frac{d\tau}{\Lambda_i^2}\exp\biggl(\tau\frac{{\cal K}}{\Lambda_i^2}\biggr).
\ee
The auxiliary propagator $\hat\Delta$ does not possess poles and does not have particles. Tree order amplitudes such as Compton scattering amplitudes are identical to their local QFT counterparts. The tree amplitudes such as Compton amplitudes are the sum of (\ref{propagatorReg}) and (\ref{auxpropagator}), and this sum gives the standard local propagator and tree graphs and they are free of unphysical couplings. The QCD and QED theories are gauge invariant, unitary and Poincar\'e invariant.

If the SM were to include a Dirac mass term:
\be
\label{Diracmass}
m\bar\psi\psi= m(\bar\psi_L\psi_R+\bar\psi_R\psi_L),
\ee
it would break the gauge invariance of the $SU(2)_L\times U(1)_Y$ electroweak sector, resulting in a nonrenormalizable electroweak theory. On the other hand, the color gluons in the QCD sector are massless, preserving the color gauge invariance of $SU(3)_c$ and its renormalizability. Therefore, in the standard model for the electroweak sector to preserve the $SU(2)_L\times U(1)_Y$ gauge symmetry and renormalizability none of the model's fermions and bosons can begin with masses, but must require them to be generated from the spontaneous symmetry breaking of $SU(2)_L\times U(1)_Y$ by the Higgs mechanism involving the scalar Higgs field~\cite{Higgs1964,Brout1964,Kibble1964,Higgs1966,Kibble1967,Halzen,Peskin,Burgess}. Initially, in the SM the charged fermions are described by Dirac fermions, formed through the link between two massless Weyl (left and right-handed) fermions and the link between them is made by the Higgs field. In the case of the electrically neutral neutrinos in the SM there is only the left-handed massless Weyl neutrino, while the right-handed massless Weyl neutrino is absent. For charged fermions and the neutrinos only the left-handed neutrinos interact with the W boson.

The gauge invariance symmetry of the Lagrangian ${\cal L}_{SMEW}$ would be broken by the mass Lagrangian terms Eq.(\ref{bosemass}) and Eq.(\ref{quarkleptonmass}). Our UCQFT model allows for an alternative interpretation of the elecroweak $SU(2)\times U(1)$ sector. Because the loop graphs are finite an infinite renormalization of particle interactions is not required. This allows for finite field theory interactions with massive bosons and fermions in the Lagrangian and finite radiative loop calculations. The idea that the $SU(2)\times U(1)$ Lagrangian has to be massless at the outset to guarantee a gauge invariant and renormalizable scheme is discarded. Moreover, the assumption that the classical Higgs potential has the form:
\be
\label{SMpotential}
{\cal L}_\phi=-\mu^2\phi^\dagger\phi+\partial_\mu\phi^\dagger\partial^\mu\phi+\lambda(\phi^\dagger\phi)^2,
\ee
where $\phi$ is the complex Higgs field need not be made. The boson and fermion masses are calculated from perturbative one-loop graphs with an associated QFT length (mass) scale $\Lambda_i$. Because the spontaneous symmetry breaking mechanism of electroweak symmetry is discarded in our model, the Higgs field vacuum expectation value $v=\langle 0|\phi|0\rangle = 0$. All the low energy predicted decay products and particle productions verified by the LHC will be retained in the alternative model. However, the only electroweak true vacuum will be $v=0$, predicting a stable vacuum in contrast to the standard prediction by the Higgs mechanism of an unstable vacuum at very high energies~\cite{Moffat2021}.

The $A_\mu$ and $Z_\mu$ are linear combinations of the two fields
$W_{3\mu}$ and $B_\mu$ that mix after the breaking of the $SU(2)_L\times U(1)_Y$ gauge symmetry by the particle mass terms and leads to states that are physically observable:
\begin{equation}
\label{A} A_\mu=c_wB_\mu+s_wW_{3\mu},
\end{equation}
\begin{equation}
\label{Z} Z_\mu=-s_wB_\mu+c_wW_{3\mu},
\end{equation}
where $c_w=\cos\theta_w$, $s_w=\sin\theta_w$ and the angle $\theta_w$ denotes the weak mixing angle. The massive charged W bosons are:
\be
W^\pm=\frac{1}{\sqrt 2}(W^1\pm iW^2).
\ee

The electroweak coupling constants $g$ and $g'$ are related to the
electric charge $e$ by the standard equation
\be
gs_w=g'c_w=e
\ee
and we use the standard normalization $c_w=g/(g^2+g^{'2})^{1/2}$ and $g'/g=\tan\theta_w$.

\section{Neutrino Sector and Masses}

In the SM only the left-handed fermions participate in the weak interactions with the $W^{\pm}$ and $Z^0$ and because no right-handed neutrino has been experimentally detected the SM neutrinos are massless and stable. Moreover, the SM asserts that the three lepton numbers $L_e$, $L_\mu$ and $L_\tau$ are conserved (up to negligible corrections involving the electroweak anomaly). It has now been well established experimentally through nuclear reactions which power the Sun, and cosmic rays detected on Earth that neutrinos are massive. These neutrino masses are significantly lighter than the other SM fermions. These experiments are not in agreement with the separate conservation of $L_e$, $L_\mu$ and $L_\tau$. The experimentally detected neutrino oscillations in which the neutrino flavors are exchanged are only possible for massive neutrinos. Because the SM fails to account for these experiments and the masses of neutrinos, it has to be extended or modified in a fundamental way.

The massless SM neutrinos can be described by left and right-handed chiral fermions corresponding to left and right helicity determined by the projection of the spin on the momentum of the particle and the opposite helicity for the antiparticle. For the massless neutrinos the weak interactions involve only left-handed spinors $\nu_L$. However, for massive fermions you can always find a Lorentz reference frame where the fermion moves in the opposite direction, so the massive fermion helicity is not a Lorentz invariant. This means that for massive fermions you need spinors of the opposite chirality, realized as the Dirac mass Eq.(\ref{Diracmass}) that mixes the spinors of opposite chirality for charged quarks and leptons. Electric charge is a Lorentz invariant, and also a constant of the motion for massive fields. The electrically neutral neutrinos are less constrained than the charged particles, because whereas chirality is a Lorentz invariant, it is not a constant of the motion for massive neutrinos. Charge conjugation defined by the operator $C$ plays an important role, as it is the symmetry that allows Majorana particles to be described as elecrically neutral. The Majorana equation is given by~\cite{Majorana}:
\be
i\slashed\psi - m\psi^c=0.
\ee
The spinor $\psi^c=C\psi$ is the charged conjugate of $\psi$. By definition
\be
\psi^c=\eta_cC\bar\psi^T,
\ee
where $(.)^T$ denotes the transpose, $\eta_c$ is an arbitrary phase factor taken conventionally as $\eta_c=1$, and $C$ is a $4\times 4$ matrix.

If a right-handed massive neutrino $\nu_R$ exists, then at low energies we would get the SM Yukawa Lagrangian contribution given by
\be
{\cal L}_Y=gv\nu_l\nu_R^\dagger +g\nu_Lh\nu_R^\dagger + h.c.,
\ee
where $h$ is the neutral Higgs particle field. The first term corresponds to the mass of the neutrino, $m_\nu=gv$, $v=\langle 0|\phi| 0\rangle = 246$ GeV is the vacuum expectation of the scalar Higgs field and  $g$ is the neutrino coupling constant. Because of the very light masses of the neutrinos compared to the mass of the electron, the coupling constant g is extremely small $g < 10^{-12}$, a fact that is difficult to understand and appears as an unnatural result in SM as it severely suppresses the coupling of neutrinos to matter.

Because the right-handed neutrino with zero quantum numbers has not been experimentally detected, we can resort to postulating that neutrinos are Majorana fermions, whereby the neutrino is the same as the antineutrino. However, in the SM to preserve the gauge symmetry $SU(2)_L\times U(1)_L$ and the coupling to the scalar Higgs field, the coupling has to be the following (for left-chirality neutrinos the following mass term changes weak hypercharge by two units, which is not possible with the standard gauge invariant Higgs interaction):
\be
\label{2Higgscoupling}
{\cal L}_{\nu h}=\frac{g}{M}(C\nu_L)^\dagger h^2\nu_L.
\ee
Here, the $C\nu=\nu^c$ denotes the charged conjugation $C\chi$ of the Majorana spinor $\chi$ that behaves as the spinor with the opposite chirality and $C\nu_L$ can play the role of the right-handed neutrino $\nu_R$. Moreover, the parameter $M$ denotes a mass scale at which the SM breaks down, loses predictability and is incomplete. In QFT the mass dimensionality of the Lagrangian is $m^p$. Renormalizable theories require $p\geq 0$, so that Eq.(\ref{2Higgscoupling}) is a nonrenomalizable interaction and is excluded by the SM. One can now resort, however, to an effective low energy QFT with added coefficients in the effective QFT Lagrangian, but this removes the possibility of obtaining a complete particle model of electroweak interactions.

In our UCQFT, we are not constrained to require that the neutrino model is $SU(2)_L\times U(1)_Y$ gauge invariant and infinitely renormalizable as the radiative correction loop graphs are finite to all orders of perturbation theory. Indeed, Eq.({\ref{2Higgscoupling}) will lead to finite loop radiative corrections. The three flavors of neutrinos labeled $\nu_e, \nu_\mu$ and $\nu_\tau$ can contribute the mass term:
\be
{\cal L}={\cal L}_{SM}+\frac{1}{2}[m_{ij}(\bar\nu_iP_L\nu_j) +h.c.],
\ee
where $m_{ij}$ is a complex symmetric matrix and $P_L=(1/2)(1-\gamma_5)$. The $\nu_i$ are left-handed neutrinos and, by assumption, the CPT takes the existing neutrinos into themselves and neutrinos must be their own antiparticles (Majorana fermions). The neutrino mass term breaks all of the lepton number symmetries of the standard model.

The neutrino mass matrix may be diagonalized by redefining the neutrino fields, $\nu_i=K_{ij}\nu_j$, where $K_{ij}$ is a unitary matrix that has fewer entries than the quark-mass matrix $q_{mn}$. After this transformation, the Lagrangian is given by
\be
{\cal L}=-\frac{1}{2}\bar\nu_i(\slashed\partial+m_i)\nu_i +{\cal L}_{\nu_i} + {\cal L}_{nc} + {\cal L}_{cc},
\ee
where the masses are real and positive. The neutral current interaction Lagrangian ${\cal L}_{nc}$ remains unchanged, because of the unitarity of $K$, while the charged current interaction ${\cal L}_{cc}$ factors in a CKM-like mixing matrix:
\be
{\cal L}_{cc}=\frac{igK_{ai}}{\sqrt{2}}W_\mu({\bar l}_a\gamma^\mu P_L\nu_i) + h.c.,
\ee
where $l_a$ denotes the charged leptons $l_1,l_2,l_3=e,\mu,\tau$. The $K_{ai}$ matrix can be expressed in terms of mixing angles and phases, $K=BA$ with
$A={\rm Diag}(\exp{(i\alpha_1/2)},\exp{(i\alpha_2/2)},\exp{(i\alpha_3/2)})$, where $A$ is the Pontekorvo, Maki, and Sakata matrix~\cite{Pontecorvo}~\cite{Maki}~\cite{Burgess}. This matrix participates in charged-current interactions with charged leptons. The phases in the A matrix will produce CP violating phases in weak interactions.

The violation of the three accidental symmetries connected to the flavour lepton numbers $L_e$,$L_\mu$ and $L_\tau$, leaving the difference between baryon number and lepton number, $B-L$, as a detectable number, can be experimentally tested in neutrinoless double beta (O$\nu\beta\beta$) decay~\cite{doublebetadecay}. This experiment can test whether neutrinos are Majorana particles.

The fermion masses in the SM are generated
through Yukawa couplings and the spontaneous symmetry breaking Higgs mechanism with $v=\langle 0|\phi| 0\rangle\neq 0$. In a previous publication~\cite{Moffat2021}, the fermion masses including the neutrino masses were generated from
the finite one-loop fermion self-energy graphs by means of a
Nambu-Jona-Lasinio mechanism~\cite{Nambu}. A fermion particle satisfies
\be
\label{meq} i\slashed{p}+m_{0f}+\Sigma(p)=0,
\ee
for $i\slashed{p}+m_f=0$ where $m_{0f}$ is the bare fermion
mass, $m_f$ is the observed fermion mass and $\Sigma(p)$ is the
finite proper self-energy part. We have
\begin{equation}
\label{sigmaeq}
m_f-m_{0f}=\Sigma(p,m_f,g,\Lambda_f)\vert_{i\slashed{p}+m_f=0}.
\end{equation}
Here, $\Lambda_f$ denote the energy scales for lepton and quark masses.

The fermion mass is identified with $\Sigma(p)$ at $p=0$ and we choose $m_{0f}=0$~\cite{Moffat2021}:
\be
m_f=\Sigma(0)=\frac{\alpha_fm_f}{\pi}\biggl[\ln\biggl(\frac{\Lambda_f^2}{m_f^2}\biggr)
-\gamma_e\biggr]
+O\biggl[\frac{\ln(\Lambda^2_f)}{\Lambda_f^2}\biggr],
\ee
where $\gamma_e$ is the Euler-Mascheroni constant, $\gamma_e=0.57722$, $\alpha_f=g_f^2/4\pi$ where $g^2_f$ is a fermion coupling constant containing quark color factors for strong coupling and is a weak coupling constant for leptons. This equation has two solutions: either $m_f=0$, or
\be
1=\frac{\alpha_f}{\pi}\biggl[\ln\biggl(\frac{\Lambda_f^2}{m_f^2}\biggr)-\gamma_e\biggr].
\ee
The first trivial solution corresponds to the standard
perturbation result. The second non-trivial solution will
determine $m_f$ in terms of $\alpha_f$ and $\Lambda_f$ and leads
to the fermion ``mass gap'' equation:
\be
m_f=\Lambda_f\exp\biggl[-\frac{1}{2}\biggl(\frac{\pi}{\alpha_f}+\gamma_e\biggr)\biggr].
\ee
The calculated fermion masses are displayed in Table 1 in~\cite{Moffat2021}.

The mass scales $\Lambda_f$ for fermions determined by Eq.(39) are well above the energies achievable by the LHC and by foreseeable future high energy accelerators. The electron mass, $m_e=0.000511$ GeV, corresponds to the mass scale $\Lambda_e=3.2\times 10^8$ TeV, while the assumed electron neutrino mass, $m_{\nu_e}=0.2\times 10^{-8}$ GeV, (the absolute neutrino masses have not yet been experimentally determined) corresponds to $\Lambda_{\nu_e}=3.2\times 10^8$ TeV. The very high energy fermion mass scales $\Lambda_f$ indicate that the nonlocal field operators smear the interaction vertices and the propagators at these ultraviolet high energy mass scales. Below these mass scales the theory behaves as a local field theory agreeing with the LHC data. However, in ref.~\cite{Moffat2021}, it was demonstrated that the W and Z boson mass scale, $\Lambda_{WZ}=542$ GeV, and the Higgs mass scale, $\Lambda_H=1.57$ TeV, are well within the experimental range of the LHC and the smearing of the interactions by the nonlocal field operators can be experimentally tested.

\section{Conclusions}

We formulated a minimal model of neutrinos in which the three flavors of massive neutrinos are described by Majorana neutrinos. The model is consistent with neutrino oscillation models and experiments. It is also possible to formulate a model incorporating a sterile neutrino, which does not interact with matter. Experiments are under way to detect Majorana neutrinos~\cite{doublebetadecay} and sterile neutrinos~\cite{sterileneutrinoexp}. The minimal neutrino model in which neutrinos are the same as their antiparticles can explain the matter-antimatter (baryon-antibaryon) asymmetry in the early universe~\cite{asymmetrymatter}. Out-of-equilibrium decays of massive Majorana neutrinos can generate a lepton asymmetry, which by sphalerons is partially transformed into a baryon asymmetry and baryongenesis.

\section*{Acknowledgments}

I thank Martin Green, Viktor Toth and Latham Boyle for helpful discussions. Research at the Perimeter Institute for Theoretical Physics is supported by the Government of Canada through industry Canada and by the Province of Ontario through the Ministry of Research and Innovation (MRI).

\end{document}